\documentclass[aps,prl,twocolumn,groupedaddress,showpacs]{revtex4-1}

\usepackage{graphicx}
\usepackage{bm}
\begin{document}

%\preprint{}

\title{Coulombic effect and renormalization in nuclear pairing}


\author{H. Nakada}
\email[E-mail:\,\,]{nakada@faculty.chiba-u.jp}
%\thanks{}
\affiliation{Department of Physics, Graduate School of Science,
 Chiba University\\
Yayoi-cho 1-33, Inage, Chiba 263-8522, Japan}

\author{M. Yamagami}
%\email[E-mail:\,\,]{}
%\thanks{}
\affiliation{Department of Computer Science and Engineering,
 University of Aizu\\
Aizu-Wakamatsu, Fukushima 965-8580, Japan}


\date{\today}

\begin{abstract}
We investigate effects of the Coulomb force on the nuclear pairing properties
by performing the Gogny Hartree-Fock-Bogolyubov calculations
for the $N=20$, $28$, $50$, $82$ and $126$ nuclei.
The Coulomb force reduces the proton pair energy
and the even-odd mass difference by about 25\%,
except for nuclei at and around the proton shell or subshell closure.
We then propose a renormalization scheme via a reduction factor $\gamma_p$
for the proton pairing channel.
It is found that a single value of $\gamma_p(=0.90)$ well takes account of
the Coulombic effect,
for nuclei covering wide range of the mass number and the neutron excess
including the nuclei around the shell or subshell closure.
\end{abstract}

%\pacs{21.30.Fe, 21.60.Jz, 21.10.Dr, 21.65.-f}
\pacs{21.60.Jz, 21.30.Fe, 21.10.Sf, 21.10.Dr}% PACS, the Physics and Astronomy
%\keywords{}

\maketitle

%\section{}
%\subsection{}
%\subsubsection{}

% If in two-column mode, this environment will change to single-column
% format so that long equations can be displayed. Use
%\begin{widetext}
%\end{widetext}

%\section{Introduction\label{sec:intro}}
\textit{Introduction.}
The energy density functional (EDF) approach (or the mean field approach)
provides us with a microscopic framework
for describing the static and dynamical properties of atomic nuclei
from the nucleonic degrees of freedom~\cite{BH03}.
Because of its numerical feasibility,
the quasi-local EDF's (\textit{i.e.} the EDF's
represented by local densities and currents
that include low-order derivatives, such as the Skyrme EDF)~\cite{DCK10}
have been applied to calculations covering wide range of the nuclear chart.
With the coordinate space representation,
the quasi-local EDF's are suitable
for describing various exotic deformations~\cite{YM01}
and continuum effects~\cite{GS01,Ma01,KS02,MM09}
of neutron-rich nuclei.

The pairing correlations play a significant role
in static and dynamic properties of nuclei at low energy~\cite{BB05,DH03}.
It is crucially important to construct the pairing channel of EDF (pair-EDF) 
reproducing the pairing properties across the nuclear chart~\cite{DN96}.
In the approaches using the quasi-local EDF's,
the pair-EDF is usually taken to be local,
in order to keep the numerical feasibility.
In most Skyrme EDF calculations so far,
the energy density of the form $\mathcal{H}_\tau^\mathrm{pair}(\mathbf{r})
= A_\tau\big[\rho(\mathbf{r})\big]\,\kappa_\tau^\ast(\mathbf{r})\,
\kappa_\tau(\mathbf{r})$ ($\tau=p,n$) has been adopted
for the pair-EDF~\cite{BH03,TG00,BB09},
where $\rho=\rho_p+\rho_n$ is the isoscalar density of nucleons
and $\kappa_\tau$ the local pair density,
by adjusting a few parameters in the function $A_\tau[\rho]$.
It is found that the strength parameter in $A_p$ is substantially stronger
than that in $A_n$ to reproduce
the observed pairing properties~\cite{TG00,BB09}.
Such asymmetry should originate in the dependence of the pairing
on the neutron excess as well as in the Coulomb force
which acts only on protons.
To include effects of the neutron excess,
$\mathcal{H}_\tau^\mathrm{pair}(\mathbf{r})$ has been extended as 
$\mathcal{H}_\tau^\mathrm{pair}(\mathbf{r})
= B_\tau\big[\rho(\mathbf{r}),\rho_1(\mathbf{r})\big]\,
\kappa_\tau^\ast(\mathbf{r})\,\kappa_\tau(\mathbf{r})$
in Refs.~\cite{MS07,YS08,YS09}, where $\rho_1=\rho_n-\rho_p$,
though keeping the charge symmetry.

Although the Coulomb force is an important ingredient of the nuclear systems,
the Coulomb force has not explicitly been included in the pair-EDF
in most systematic calculations because of its non-local nature.
It was reported that the proton pairing gaps are reduced by $20-30\%$
if the Coulomb repulsion is treated self-consistently~\cite{AE01,LD09}.
It is not likely that the charge symmetric pair-EDF
appropriately represents the Coulombic effect.
An approximate method to take into account
the Coulombic effect with keeping the local nature
could be renormalizing the strength parameter of the proton pair-EDF
as in Ref.~\cite{CG08}.
However, it is not obvious whether such a simple renormalization scheme
works sufficiently well.
Moreover, an appropriate value of the renormalization parameter
and its dependence on $Z$ and $N$ have not been known.

In this paper we investigate Coulombic effect on the nuclear pairing
by the self-consistent Hartree-Fock-Bogolyubov (HFB) calculations,
particularly focusing on the renormalizability.
A numerical method that is applicable to wide range of the nuclear chart
with a finite-range interaction is required for this purpose.
Note that the HFB theory with a finite-range interaction
is practically identical to the approach with a non-local EDF.
We employ the Gaussian expansion method~\cite{NS02,Nak06},
which is adaptable to drip-line nuclei even with finite-range interactions.
For both the particle-hole (ph) and particle-particle (pp) channels,
we adopt the Gogny-D1S~\cite{D1S} plus Coulomb interaction
with the center-of-mass correction.
Although we restrict ourselves
to the $N=20$, $28$, $50$, $82$ and $126$ nuclei,
assuming the spherical symmetry,
they distribute over wide range of the mass number $(30\leq A\leq 220)$
and the neutron excess $(-0.13\leq (N-Z)/A\leq 0.36)$.
It is numerically examined whether the Coulombic effect on the pairing
can be incorporated by a renormalization factor for the proton pairing.

%\section{Hamiltonian\label{sec:Hamil}}
\textit{Hamiltonian.}
We here describe the EDF in terms of the effective Hamiltonian.
The Hamiltonian for the HFB calculations consists of
the nuclear part and the Coulomb interaction,
\begin{equation} H = H^N + V^C\,,
\end{equation}
where $H^N = K + V^N - H^\mathrm{c.m.}$
with the kinetic energy $K$, the effective nuclear interaction $V^N$,
and the center-of-mass Hamiltonian $H^\mathrm{c.m.}$.
$V^N$ may include many-body forces,
which are often simplified by a density-dependent two-body force.
The HFB energy can be represented in the EDF form
owing to Wick's theorem,
though including non-local terms in general.
For $V^N$ we adopt the D1S parameter-set of the Gogny interaction
in this paper,
which can be employed without the energy cut-off for the pairing channel.
Since the short-range $NN$ correlation hardly influences
matrix elements of $V^C$,
we use the bare Coulomb force for $V^C$.
The spherical HFB calculations are implemented
for the $N=20$, $28$, $50$, $82$ and $126$ nuclei,
by applying the Gaussian expansion method~\cite{NS02,Nak06},
with the basis functions given in Ref.~\cite{Nak08}.
It is noted that the exchange term of $V^C$ is treated exactly,
and that both one- and two-body terms of $H^\mathrm{c.m.}$
are subtracted before iteration.

In correspondence to the expression of the HFB energy
by the density matrix and the pairing tensor~\cite{RS80},
we separate the Hamiltonian
into the pp part $H_\mathrm{pp}$ that gives the pairing tensor
and the ph part $H_\mathrm{ph}$.
Each of them is comprised of the nuclear and the Coulomb parts;
\begin{equation}
 H_\mathrm{ph} = H^N_\mathrm{ph} + V^C_\mathrm{ph}\,,\quad
 H_\mathrm{pp} = H^N_\mathrm{pp} + V^C_\mathrm{pp}\,.
\end{equation}
We consider the pairing between like nucleons as usual,
neglecting the proton-neutron pairing,
which is not important except $Z\approx N$ cases.
$H^N_\mathrm{pp}$ is then separable into the proton and neutron parts,
\begin{equation}
 H^N_\mathrm{pp} = H^p_\mathrm{pp} + H^n_\mathrm{pp}\,.
\end{equation}
The proton pairing should be subject
to $H^p_\mathrm{pp}+V^C_\mathrm{pp}$.
If the Hamiltonian contains only the zero-range interactions,
we need only the local limit
of the density matrix and the pairing tensor,
which leads to a local or quasi-local EDF.
However, the interactions have finite range in general,
and it is not obvious whether and how the energy of nuclei can be
approximated to sufficient precision by the local limit.
In particular,
whereas there have been validating arguments
for $H_\mathrm{ph}$~\cite{DCK10,NV72},
local approximation for $H_\mathrm{pp}$
from which $\mathcal{H}_\tau^\mathrm{pair}(\mathbf{r})$ is derived
has not been well explored.

In the HFB calculations of nuclei,
we reasonably postulate that $H^N$ is isoscalar.
Acting only on protons, $V^C$ breaks the charge symmetry.
While the charge symmetry is broken at the Hartree-Fock level,
there should also be difference between $H_\mathrm{pp}$ for protons
and for neutrons because of $V^C_\mathrm{pp}$. 
$H^N_\mathrm{pp}$ has often been determined
so as to reproduce the observed pairing properties
among neutrons~\cite{MS07,YS08,D1S,Nak08b,Nak10};
namely, by using only $H^n_\mathrm{pp}$.
To examine whether effects of $V^C_\mathrm{pp}$ can be treated
in a simple renormalization scheme,
we define the following Hamiltonian,
\begin{equation}
 \bar{H}(\gamma_p) = H^N_\mathrm{ph} + V^C_\mathrm{ph}
 + \gamma_p\,H^p_\mathrm{pp} + H^n_\mathrm{pp}
 = (H - V^C_\mathrm{pp}) - (1-\gamma_p)\,H^p_\mathrm{pp}\,,
\end{equation}
dropping $V^C_\mathrm{pp}$ and introducing the renormalization parameter
$\gamma_p$.
While the charge symmetry in the pairing channel does not hold
because of $V^C_\mathrm{pp}$ in $H$,
$-(1-\gamma_p)\,H^p_\mathrm{pp}$ gives the charge symmetry breaking
in the pairing channel of $\bar{H}(\gamma_p)$.
Note that many HFB calculations so far have employed $\bar{H}(1)$,
by presuming the charge symmetry for $H_\mathrm{pp}$.
The central question here is whether or not we have
\begin{equation}
 \langle H\rangle_H \approx
 \langle \bar{H}(\gamma_p)\rangle_{\bar{H}(\gamma_p)}\,,
 \label{eq:approx1}
\end{equation}
with an appropriate $\gamma_p$.
We can then renormalize $H^N_\mathrm{pp}$ via $\gamma_p$
to represent the Coulombic effect.
The prescription for the Coulombic effect using $\gamma_p$
has been applied to $^{17}$Ne within a three-body model~\cite{OHS10}.
Since we here carry out self-consistent HFB calculations,
the HFB energy at the left-hand side (lhs) is evaluated
by a calculation with the full Hamiltonian $H$,
while the energy at the right-hand side (rhs) with $\bar{H}(\gamma_p)$.
These Hamiltonians are explicitly written as subscripts
in Eq.~(\ref{eq:approx1}).
If the wave functions are similar,
we have $\langle H^N_\mathrm{ph} + V^C_\mathrm{ph} + H^n_\mathrm{pp}\rangle_H
\approx \langle H^N_\mathrm{ph} + V^C_\mathrm{ph} + H^n_\mathrm{pp}
\rangle_{\bar{H}(\gamma_p)}$
and Eq.~(\ref{eq:approx1}) therefore indicates
\begin{equation}
 \langle H^p_\mathrm{pp}+V^C_\mathrm{pp}\rangle_H \approx
 \langle \gamma_p\,H^p_\mathrm{pp}\rangle_{\bar{H}(\gamma_p)}\,,
 \label{eq:approx2}
\end{equation}
which is further reduced to
\begin{equation}
 \langle V^C_\mathrm{pp}\rangle_H \approx
 -\langle (1-\gamma_p)\,H^p_\mathrm{pp}
 \rangle_{\bar{H}(\gamma_p)}\,, \label{eq:approx3}
\end{equation}
via $\langle H^p_\mathrm{pp}\rangle_H \approx
\langle H^p_\mathrm{pp}\rangle_{\bar{H}(\gamma_p)}$.
The value of $\gamma_p$ may be determined
for individual nucleus.
However, for the renormalization scheme via $\gamma_p$ to be useful,
$\gamma_p$ has to be fixed without referring the result of $H$
for individual nucleus.
It is hence desired that $\gamma_p$ is insensitive to nuclide
or expressed by a simple function of $Z$ and $N$.
In this work we consider the simplest case that $\gamma_p$ is a constant,
with no $Z$ or $N$ dependence.

\begin{figure}
%\begin{figure*}
%\includegraphics[scale=0.75]{Ep-all.eps}
\includegraphics[scale=0.36]{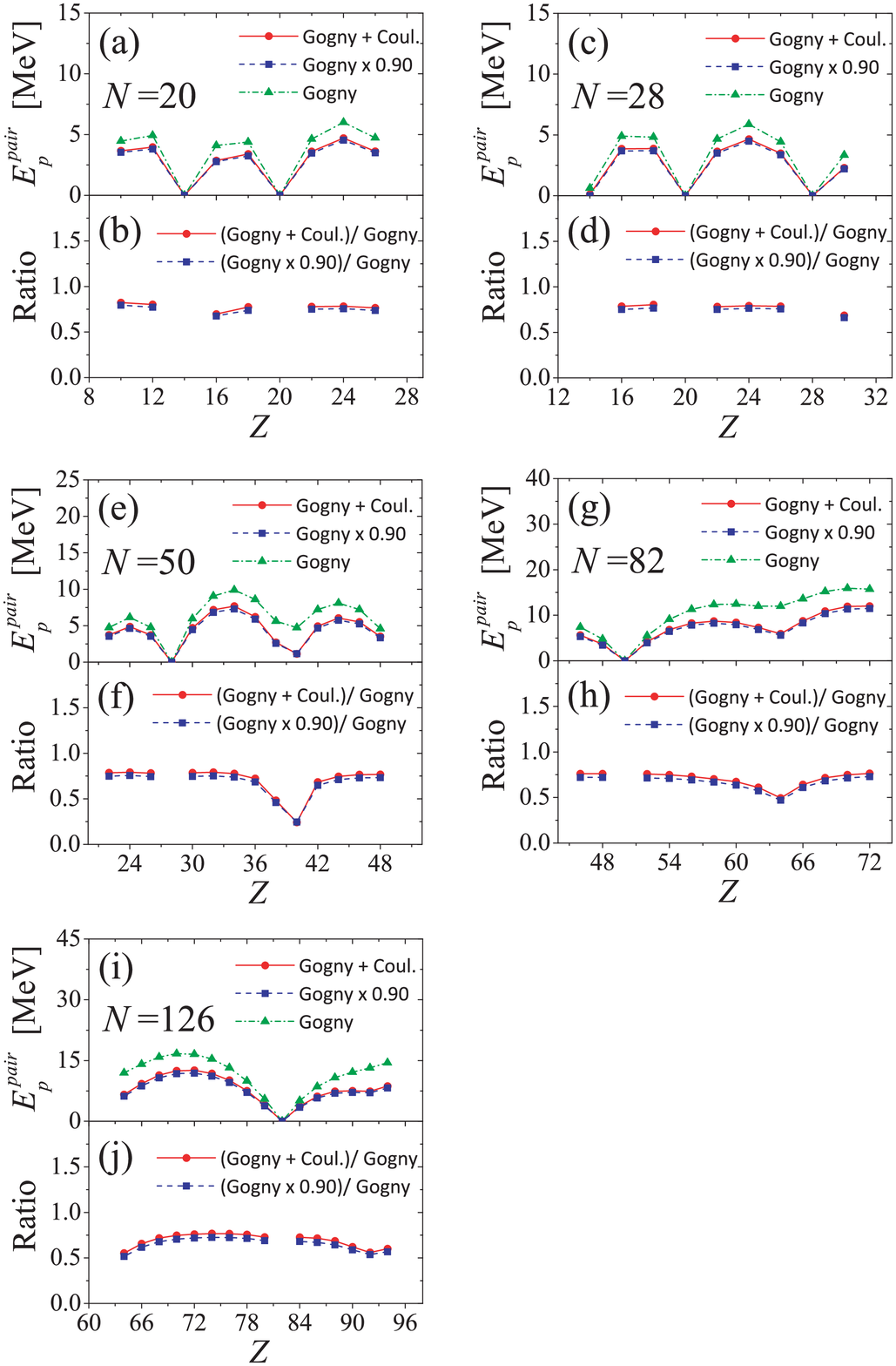}
\caption{(Color) Comparison of the pair energy $E_p^\mathrm{pair}$
 obtained from the HFB calculations:
 (a) $E_p^\mathrm{pair}$ obtained with $H$ (red circles),
 $\bar{H}(1)$ (green triangles) and $\bar{H}(0.90)$ (blue squares),
 (b) ratios of $E_p^\mathrm{pair}$ obtained with $H$ (red circles)
 and $\bar{H}(0.90)$ (blue squares) relative to
 those obtained with $\bar{H}(1)$,
 for the $N=20$ isotones.
 Analogously, $E_p^\mathrm{pair}$ and ratio for the (c,d) $N=28$,
 (e,f) $N=50$, (g,h) $N=82$, and (i,j) $N=126$ isotones.
 Lines are drawn to guide eyes.
\label{fig:Epair}}
%\end{figure*}
\end{figure}

\begin{figure}
%\begin{figure*}
%\includegraphics[scale=0.75]{DEL-all.eps}
\includegraphics[scale=0.36]{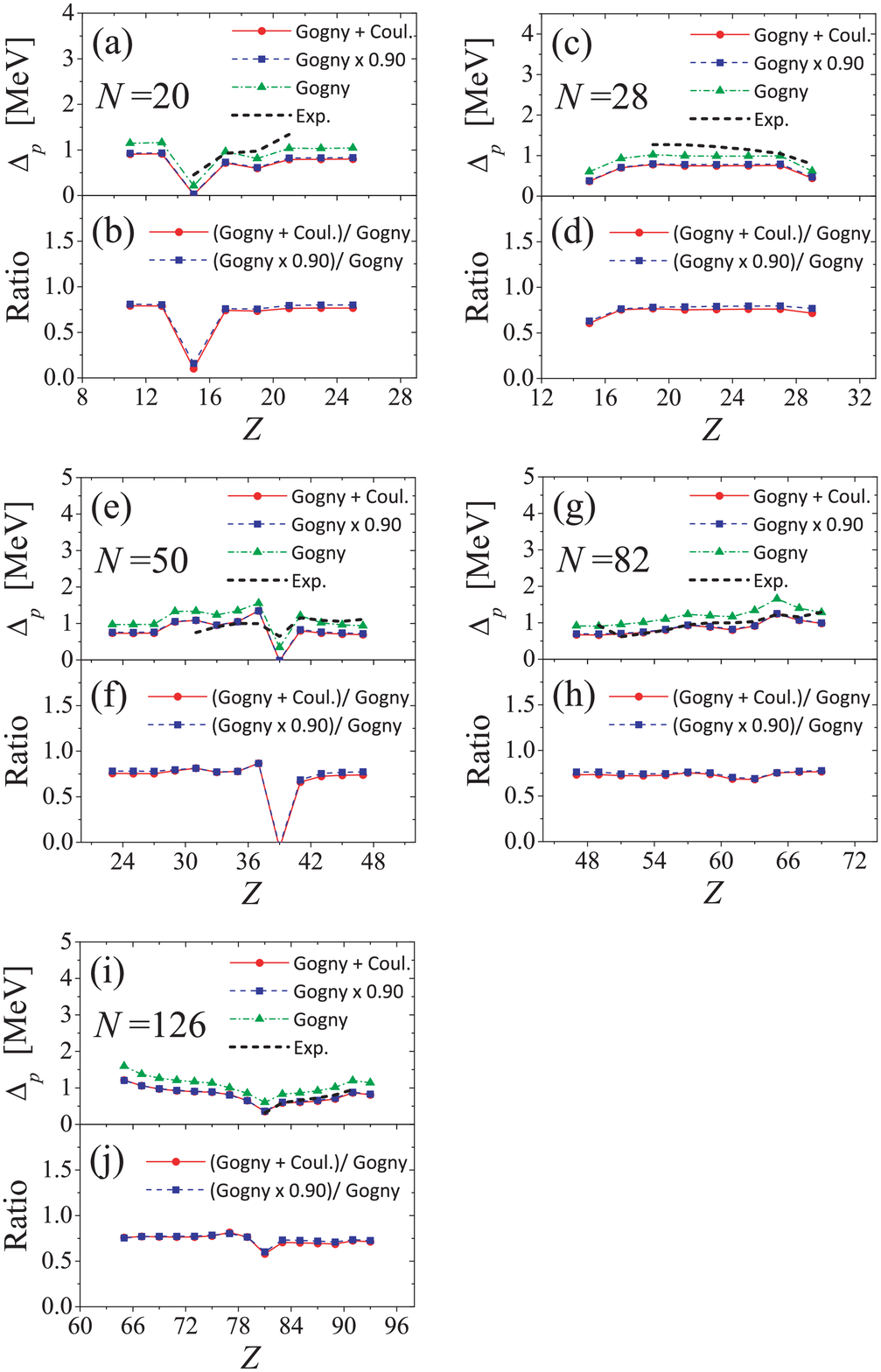}
\caption{(Color) Comparison of the even-odd mass difference $\Delta_p$
 and of its ratio obtained from the HFB calculations
 for the (a,b) $N=20$, (c,d) $N=28$, (e,f) $N=50$, (g,h) $N=82$,
 and (i,j) $N=126$ isotones.
 See Fig.~\protect\ref{fig:Epair} for conventions.
 Experimental values taken from Ref.~\protect\cite{mass} are shown
 by black short-dashed lines in (a,c,e,g,i).
\label{fig:massdif}}
%\end{figure*}
\end{figure}

%\section{Results\label{sec:result}}
\textit{Results.}
Figures~\ref{fig:Epair} and \ref{fig:massdif} depict
the spherical HFB results of the pair energy (for $Z=\mathrm{even}$ nuclei)
and the even-odd mass difference (for $Z=\mathrm{odd}$ nuclei)
in the $N=20$, $28$, $50$, $82$ and $126$ isotones~\cite{basis}.
Because the neutron pair energy vanishes in these calculations,
the even-odd mass difference is straightforwardly connected
to the proton pairing.
While several neutron-rich $N=20$ and $28$ nuclei seem to be deformed
in reality~\cite{SP08},
it is sufficiently meaningful to compare the results
of $H$ and $\bar{H}(\gamma_p)$ within the spherical HFB calculations.
The pair energy $E_p^\mathrm{pair}(=E^\mathrm{pair})$ is defined
by the energy contribution of $H_\mathrm{pp}$
(\textit{i.e.} $\langle H^p_\mathrm{pp}+V^C_\mathrm{pp}\rangle_H$ or
$\langle \gamma_p\,H^p_\mathrm{pp}\rangle_{\bar{H}(\gamma_p)}$),
which is a simple and clear indicator to the pairing.
The even-odd mass difference is defined by
\begin{equation}
 \Delta_p(Z) = E(Z,N) - \frac{1}{2}\big[E(Z-1,N) + E(Z+1,N)\big]\,,
\end{equation}
for $Z=\mathrm{odd}$ nuclei.
The HFB energies of the $Z=\mathrm{odd}$ nuclei
are calculated in the equal-filling approximation~\cite{EFA},
which has been shown to work well~\cite{EFAb}.
As an observable directly corresponding to the data,
$\Delta_p$ has clear physical meaning.

Let us first compare the $E_p^\mathrm{pair}$ and $\Delta_p$ values
without $V^C_\mathrm{pp}$ (\textit{i.e.} by $\bar{H}(1)$,
green triangles in Figs.~\ref{fig:Epair} and \ref{fig:massdif})
and those of the full Hamiltonian $H$ (red circles).
Similar comparison was made in Ref.~\cite{LD09},
although they viewed the pairing gap of the canonical basis
locating adjacent to the Fermi energy.
It is found that, both for $E_p^\mathrm{pair}$ and $\Delta_p$,
the ratio of the value of $H$ to that of $\bar{H}(1)$ is about $75\%$
(Fig.~1 (b,d,f,h,j) and Fig.~2 (b,d,f,h,j)).
This result seems consistent with those in Refs.~\cite{AE01,LD09}.
This ratio is almost stable for the nuclides under consideration
except those in the vicinity of $^{34,42}$Si, $^{90}$Zr, $^{146,190}$Gd,
$^{208}$Pb and $^{218}$U.
$^{208}$Pb is a typical doubly-magic nucleus.
The $^{90}$Zr and $^{146}$Gd nuclei are well-known
as the proton-subshell-closed ones,
reasonably having suppressed $E_p^\mathrm{pair}$ and $\Delta_p$.
Similar suppression takes place in $^{34,42}$Si, $^{190}$Gd and $^{218}$U
in the spherical HFB calculation with the D1S interaction,
whereas $^{42}$Si has been suggested to be deformed
by experiments~\cite{Si42}.
Because of the subshell nature,
the ground states of these nuclei lie around the boundary
between the normal fluid and the superfluid phases.
Hence the usually perturbative $V^C_\mathrm{pp}$ affects
$E_p^\mathrm{pair}$ and $\Delta_p$ to exceptional extent.
The same consequence was reported in Ref.~\cite{AE01} for $^{90}$Zr.

We next apply the Hamiltonian $\bar{H}(\gamma_p)$
to the self-consistent HFB calculations,
adjusting $\gamma_p$ so as to reproduce $E_p^\mathrm{pair}$ and $\Delta_p$
obtained from the full Hamiltonian $H$.
We find that a single value $\gamma_p=0.90$
satisfies Eq.~(\ref{eq:approx2}) to good approximation
all over the nuclei in this wide range of $A$ and $(N-Z)/A$,
as is clear by comparing the blue squares with the red circles
in Figs.~\ref{fig:Epair} and \ref{fig:massdif}.
Remark that this is true even for the nuclei around $^{34,42}$Si,
$^{90}$Zr, $^{146.190}$Gd, $^{208}$Pb and $^{218}$U,
in which the Coulombic effect looks exceptionally strong.
It has been confirmed that the difference in the HFB energies
between $H$ and $\bar{H}(0.90)$ is less than $0.1\,\mathrm{MeV}$,
indicating that Eq.~(\ref{eq:approx1}) itself is fulfilled
to good precision.
As the wave functions of $\bar{H}(0.90)$ resemble those of $H$,
Eq.~(\ref{eq:approx3}) with $\gamma_p=0.90$ is good as well.
Thus the full Hamiltonian $H$ is well approximated
by the renormalized Hamiltonian $\bar{H}(0.90)$ in the HFB calculations,
from stable to unstable nuclei in wide mass range.

The purpose of the present work is to investigate the Coulombic effect
in the pairing channel within the HFB framework.
However, we also display the experimental data of $\Delta_p$ for reference
in Fig.~\ref{fig:massdif},
excluding those of the neutron-rich $N=20$ and $28$ nuclei
that have been indicated to be deformed.
The HFB results of $H$ and $\bar{H}(0.90)$ are
in better agreement with the data than those of $\bar{H}(1)$
for the $N=82$ and $126$ nuclei,
but not for the $N=20$ and $28$ nuclei.
This might imply a room to improve the pairing channel in the D1S interaction
or influence of additional correlations.

In the present work all the calculations are implemented
by using the Gogny-D1S interaction.
However, we have confirmed that Eq.~(\ref{eq:approx3}),
in which the lhs is the long-range Coulomb interaction
while the rhs is a short-range nuclear interaction,
is well fulfilled for wide range of mass region.
This indicates that suppression of the pair correlation
by such weak repulsion is not sensitive to interaction form.
Therefore the renormalization of the pairing channel with $\gamma_p$
will plausibly be applicable to other interactions or pair-EDF's
including local pair-EDF's.
Moreover, while the value of $\gamma_p$ may somewhat depend on
$H^N_\mathrm{pp}$ or the pair-EDF,
it should not largely deviate from $0.90$
as far as the pairing has appropriate strength.
Although charge-symmetric pair-EDF's have been assumed
in the usual Skyrme EDF approaches~\cite{BH03,MS07,YS08,YS09},
the charge symmetry in the pair-EDF should be broken
because of $V^C_\mathrm{pp}$.
It is desirable to readjust the pair-EDF (with $\rho$ and $\rho_1$ dependence)
by taking into account this Coulombic effect,
for which the renormalization with $\gamma_p(\approx 0.90)$ will be useful.

%\section{Summary\label{sec:summary}}
\textit{Summary.}
We have investigated influence of the Coulomb interaction
on the pairing channel in the spherical HFB calculations.
Using the Gogny-D1S plus Coulomb interaction
for the neutron-closed nuclei of $N=20$, $28$, $50$, $82$ and $126$,
we have found that the Coulomb interaction reduces
the pair energy and the even-odd mass difference by about 25\%,
%compared to the results without the Coulomb force,
except for nuclei around the proton-shell- or subshell-closed ones
$^{34,42}$Si, $^{90}$Zr, $^{146.190}$Gd, $^{208}$Pb and $^{218}$U.
Because of the non-local nature,
explicit inclusion of the Coulomb force is not adaptable
to the local or quasi-local EDF approaches.
As a renormalization scheme,
we have introduced a reduction factor for the proton pairing channel
of the nuclear force (or the pair-EDF),
and adjusted the factor to the results with the Coulomb interaction.
It is found that the Coulombic effect is approximated
with a single renormalization factor $\gamma_p(=0.90)$ to good precision,
all over the nuclei under consideration ranging $30\leq A\leq 220$
and $-0.13\leq (N-Z)/A\leq 0.36$,
even including the shell- or subshell-closed nuclei.

In the present work we have numerically investigated
the Coulombic effect on the pairing
and the renormalizability with $\bar{H}(\gamma_p)$.
It is of interest to justify the renormalization scheme
from a microscopic viewpoint,
though it is left as a future work.

\begin{acknowledgments}
This work is financially supported
as Grant-in-Aid for Scientific Research (C), No.~22540266,
and as Core-to-Core Program
``International Research Network for Exotic Femto Systems (EFES)'',
by Japan Society for the Promotion of Science.
Numerical calculations have been performed
on HITAC SR11000 at IMIT, Chiba University,
HITAC SR11000 at IIC, Hokkaido University,
and Altix3700 BX2 at YITP, Kyoto University.
\end{acknowledgments}


\begin{thebibliography}{99}
\bibitem{basis} For the calculation of the $N=126$ nuclei,
 we adopt the $\ell\leq 8$ single-particle space
 instead of $\ell\leq 7$~\cite{Nak08}.
 Influence of the $\ell=8$ states is confirmed to be insignificant
 for the lighter nuclei.
\end{thebibliography}

\end{document}